\documentclass[10pt,table]{wlscirep}

\usepackage{xcolor}
\usepackage[utf8]{inputenc}
\usepackage[T1]{fontenc}
\usepackage{tabularx,array,booktabs,multirow}
\usepackage{upgreek}
\usepackage{siunitx}
\usepackage{tabu}
\DeclareSIUnit\angstrom{\text{Å}}
\DeclareSIUnit\torr{\text{Torr}}

\renewcommand{\thesection}{\Roman{section}} 
\renewcommand{\thesubsection}{\thesection.\Roman{subsection}}
\newcolumntype{C}{>{\Centering\hspace{0pt}}X}

\newcommand{\minus}{\scalebox{0.50}[1.0]{$-$}}

\title{Low-loss silicon nitride Kerr-microresonators fabricated with metallic etch masks via metal lift-off}

\author[1,2,*]{Gabriel M. Colaci\'{o}n}
\author[1,2]{Lala Rukh}
\author[1,2]{Franco Buck}
\author[1,2,3,*]{Tara E. Drake}
\affil[1]{Optical Science \& Engineering Program, University of New Mexico, Albuquerque, NM 87106, USA}
\affil[2]{Center for High Technology Materials, University of New Mexico, Albuquerque, NM 87106, USA}
\affil[3]{Department of Physics \& Astronomy, University of New Mexico, Albuquerque, NM 87106, USA}

\affil[*]{gcolacion@unm.edu, drakete@unm.edu}

\def\SiN{Si$_{\text{3}}$N$_{\text{4}}$}
\def\SiO{SiO$_{\text{2}}$}
\def\LiNBO{LiNbO$_{\text{3}}$}
\def\CFetch{CF$_{\text{4}}$}
\def\oxygen{O$_{\text{2}}$}
\def\Qloaded{Q_{loaded}}
\def\Qint{Q_{int}}
\def\Qext{Q_{ext}}

\keywords{silicon nitride, frequency combs, photonic integrated circuits, Kerr-microresonators, microfabrication, etching}

\begin{abstract}
Stoichiometric silicon nitride has emerged as a widely used integrated photonic material owing to its high index of refraction, nonlinear optical properties, and broad transparency window spanning visible to mid-IR frequencies. However, silicon nitride is generally more resistant to reactive ion etching than are typical etch masks made of polymer-based resist.
This necessitates resist layers that are significantly thicker than the silicon nitride and results in mask patterns which are tall and narrow. These high-aspect-ratio patterns inhibit the plasma transport of reactive ion etching, which leads to difficulties in accurately reproducing dimensions and creating well-defined, vertical waveguide sidewalls. In this work, we overcome these challenges by developing a metallic etch mask deposited via metal lift-off that provides a $30:1$ nitride-to-metal etch rate ratio, representing a near 45-fold reduction in the required mask thickness. We demonstrate the validity of this technique by etching microring resonators with near-vertical waveguide sidewalls and intrinsic quality factors of over 1 million. Leveraging the low optical loss of our resonators, we generate optical frequency combs with more than an octave of bandwidth and dual dispersive waves. These results establish metal lift-off as a viable and easy-to-implement technique capable of producing low optical loss waveguides.
\end{abstract}

\begin{document}

\flushbottom
\maketitle

\thispagestyle{empty}


\section{Introduction}

Developing photonic integrated circuits to replace tabletop optics analogs represents a compelling new direction in many sectors of scientific research, including next-generation timing and navigation\cite{Newman19, Ropp23, Blumenthal2024}, optical computation\cite{Feldmann21,Zhu22,Wu24}, and chip-scale quantum computers\cite{OBrien09, Kues17, Rad25}. Similar to the integrated circuit (IC) revolution of the 20th century, the past two decades have seen rapid innovation in photonic integrated circuits (PICs), often aided by existing IC fabrication expertise. However, unlike their electronic antecedents, bulk commercial availability of PICs is in its infancy\cite{Shekhar2024}. In order to move this technology from the lab to the market, optical scientists and engineers often must adapt specialized processes developed in academic cleanrooms for the standard tools of IC- and MEMS-based commercial-scale foundries\cite{Siew2021,Zang2024}.

A good illustration of the difficulty of translating PIC fabrication to foundry processes is the on-chip waveguide. Waveguides are ubiquitous in PIC design as the main way of porting photons from device to device. Waveguides can also be formed into resonators\cite{Bogaerts2012,Gaeta2019}, couplers and splitters \cite{Gallacher2022}, dichroic elements\cite{Huffman2018,Dong2016}, gratings\cite{Dai2011,Gatkine2017}, and electro-optic components \cite{Zhang2021}. They are generally made of a material that is transparent within the spectrum of interest and are thus typically formed in dielectrics with large band gaps. A typical scheme for waveguide fabrication involves depositing a layer of photo- or electron resist on top of a layer of the dielectric, lithographically defining a pattern in the resist to serve as a protective mask, and removing the dielectric not protected by the mask through reactive ion etching (RIE)\cite{Li2013,Li2017,Ye2019,Puckett2021,Lu2022}.

However, commonly used dielectrics such as silica (\SiO), silicon nitride (\SiN), aluminum nitride (AlN), and lithium niobate (\LiNBO) are resistant to the physical and chemical etching of RIE. In fact, the ratio of the etching rate of the dielectric compared to that of the resist mask, the etch selectivity, is often less than $\mathrm{1:1}$\cite{ElDirani2019}. This creates challenges for fabrication, as the protective masks must be thicker than the dielectric to outlast the etching. These thick masks with tall, narrow features exhibit inhibited plasma transport and increased etch by-product redeposition as a function of feature size\cite{Gottscho1992}.
This can lead to non-uniform etching rates between features of varying sizes, sidewalls with non-vertical and/or bowed profiles, and deviations in device dimensions from the intended design. For applications that require deterministic control of properties that depend on accurate geometries and dimensions, such as resonator dispersion and coupler efficiency, etch masks of thick resist represent a significant problem. These complications can be partially alleviated with harder, more resistant masks such as hydrogen silsesquioxane (HSQ), but post-etch removal of HSQ necessitates the use of strong acids which will also remove \SiN\cite{Xuan2016}. An alternative approach to subtractive fabrication consists of depositing \SiN\,into pre-etched \SiO\,trenches; this method is particularly suited to creating thick waveguides while avoiding tensile stress and cracking. However, the \SiO\,preform etching is still subtractive, and thus the problems of sloped sidewalls and inaccurate dimensions persist\cite{Epping2015,Pfeiffer2016}.

In this work, we develop a process for the subtractive fabrication of silicon nitride microring waveguides using a metal-lift-off-based mask and subsequent subtractive etching. Metal lift-off is a common MEMS process that can be found in most commercial foundries. Metal lift-off's suitability for patterning small features in a scalable manner makes it a particularly good candidate for patterning photonic etch masks\cite{Lim2011,Aryal2022}. The metal mask in our work demonstrates improved etch selectivity of $\mathrm{30:1}$, allowing thin layers of metal to protect and template thick layers of silicon nitride. Our results show vertical sidewall angles, minimal RIE lag, and predictable etch dimensions. We optically characterize ring resonators to demonstrate reasonably low optical loss and good resonator quality factors. Furthermore, we investigate our leading sources of loss and identify paths towards improvement in the case when ultra-low loss is necessary. Finally, we demonstrate the viability of our process by generating optical frequency combs with more than an octave of bandwidth, as well as creating long-lived optical soliton states in our microring resonators.


\section{Fabrication Process}

The waveguides under study are fabricated on chiplets cut from stoichiometric \SiN\,(hereafter SiN) deposited in-house on oxidized silicon wafers. The SiN layer is deposited on top of \SI{3}{\micro\meter} of \SiO\,via low-pressure chemical vapor deposition (LPCVD) in a two-stage deposition process for a total thickness of \SI{620}{\nm}.
\begin{figure}[!htb]
\centering
\includegraphics[width=\linewidth]{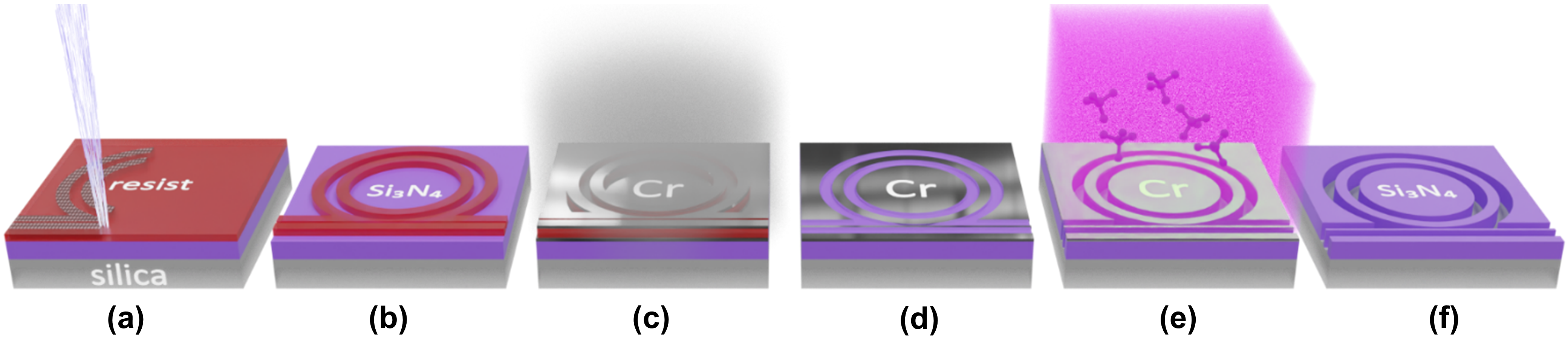}
\caption{Illustration of the process for fabricating waveguides using metal lift-off. (a) Electron-beam lithography (EBL). The process begins by coating the SiN sample with resist and writing the inverse pattern using EBL. (b) Resist development. After developing the patterned resist, the exposed (negative-tone) resist is left behind. (c) Metal deposition. A uniform layer of metal is deposited on the surface via electron beam evaporation. (d) Metal lift-off. A solvent soak/rinse removes the resist and overlaying metal to leave behind a protective etch mask. (e) ICP-RIE etching. The etch mask pattern is transferred to the SiN. (f) Mask removal. The final device is a SiN microring resonator with evanescent coupling to a bus waveguide. (\CFetch\,crystal structure in (e) is modified from PubChem database\cite{PubChem}.)}
\label{fig:fabrication flow}
\end{figure}
The full fabrication flow from unpatterned SiN chiplet to final optical device is illustrated in Fig. \ref{fig:fabrication flow}. We start with electron-beam lithography (EBL) using a JEOL JBX6300-FS to write the negative of our waveguide design in a layer ma-N 2403 negative-tone EBL resist. We employ negative-tone resist as it allows us to reduce the required EBL write area with the added benefit of improved thermal stability for deposition of the metallic mask. A \SI{50}{\nm} blanket layer of chromium (Cr) is deposited over top of the developed pattern via electron-beam physical vapor deposition (EBPVD) for a $\mathrm{5:1}$ resist to metal ratio. Lift-off is achieved with a 12-hour soak in N-Methylpyrrolidone (NMP) followed by an acetone and isopropyl alcohol (IPA) rinse. After lift-off, etching is performed using an inductively coupled plasma-reactive ion etch (ICP-RIE) tool (PlasmaTherm SLR) with a \CFetch/\oxygen/Ar chemistry. We achieve a SiN etch rate of about \SI{210}{\nm\per\min} compared to an etch rate of about \SI{7}{\nm\per\min} for Cr, for an estimated etch selectivity of $\mathrm{30:1}$. Following the etch, the remaining chromium mask is removed in a wet chromium etchant (CR-7). In order to facilitate edge-on optical coupling to the bus waveguides, a final etch is performed through the edges of the sample to form the coupling facets.
\begin{figure}[!htb]
\centering
\includegraphics[scale=1]{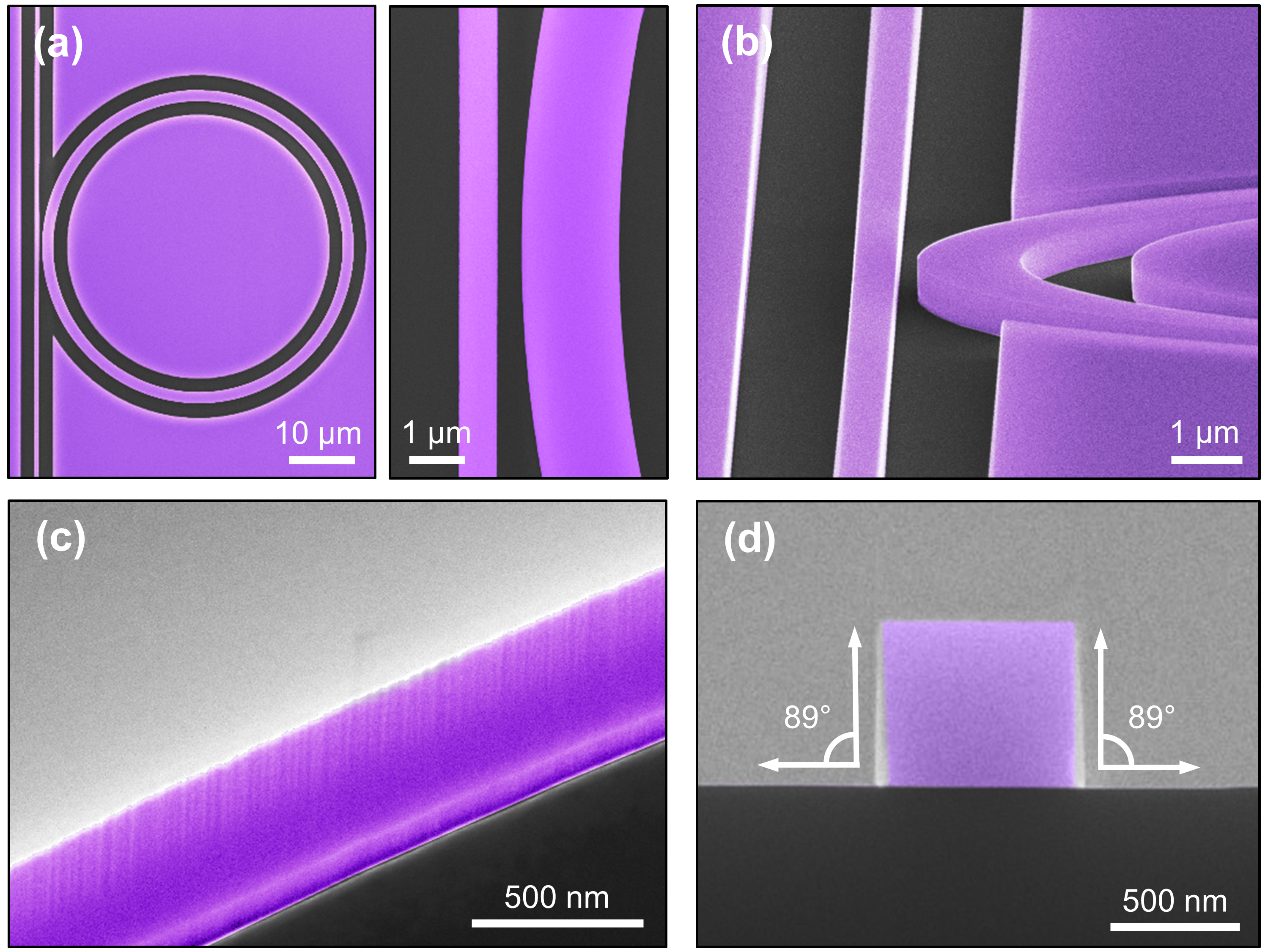}
\caption{SEM micrographs of etched waveguides with false coloring of SiN (purple) and \SiO\, (dark gray). (a) Top-down view of a microring resonator and bus waveguide (left) with zoomed in view of coupling gap region (right). (b) Angled view looking through the coupling region. (c) High magnification SEM of etched ring resonator sidewall roughness and remaining Cr mask overtop (silver). 
(d) Cleaved cross section of the coupling waveguide showing near-vertical sidewall angles. (Charging during SEM imaging adds some uncertainty to our angle measurement.)}
\label{fig:SEMbest}
\end{figure}

The SEM micrographs in Fig. \ref{fig:SEMbest} show a typical microring resonator fabricated with our metallic mask process. The left panel of Fig. \ref{fig:SEMbest}a shows a top-down view of the devices under study which consist of a \SI{46.6}{\micro\meter} diameter ring resonator and coupling (or “bus”) waveguide etched in \SI{620}{\nm} of SiN. The right panel of Fig. \ref{fig:SEMbest}a shows a higher magnification view of the coupling region where light evanescently couples across a small air gap into- and out-of- the ring. Fig. \ref{fig:SEMbest}b shows an angled view looking through this coupling region in order to highlight the clean and uniform etching of the small coupling gap (roughly \SI{400}{\nm} wide for the device shown). Some sidewall roughness, presumably transferred from the metallic mask during the etch, is evident in the ring sidewall in Fig. \ref{fig:SEMbest}c. The thin chromium mask results in near-vertical waveguide sidewall angles as shown in the cleaved cross section of the bus waveguide in Fig. \ref{fig:SEMbest}d.


\section{Optical Loss} \label{sec:Optical Loss}

Optical losses in photonic waveguides, specifically propagation losses, are classified into two sub-groups. First is loss due to optical absorption of the propagating photons by the waveguide material. SiN deposition by LPCVD produces films with residual N-H bonds, which increase absorptive losses in the telecom C-band. We minimize absorption due to N-H bonds with a high-temperature post-deposition anneal\cite{Ji2021}. The second type of propagation loss is connected to waveguide geometry and includes scattering at the core/cladding interface (sidewall scattering) and reduction of total internal reflection due to tight bending. These two effects are linked, as a tighter waveguide bending radius will cause increased overlap between the optical mode and the sidewall, enhancing sidewall scattering. Loss due to sidewall scattering is commonly attributed to fabrication-induced roughness, often originating from an uneven etch mask edge. In our work, the smoothness of the etch mask is primarily determined by two process steps, e-beam lithography and metal deposition/lift-off. Pattern roughness in electron-beam lithography is related to the pixelation of the pattern to the machine's grid. Both the diameter of the electron beam (controlled by the beam current) and the pixel separation (shot pitch) can influence the roughness of the resist sidewall based on the degree of overlap. Since the resist serves as the template for metal deposition, this roughness can be transferred to the metal mask edge. The metal mask may acquire additional roughness based on the details of the EBPVD step, including the chamber base pressure and the evaporation rate of the metal. 

In order to investigate the optical losses induced by our process, we pattern a series of waveguide-coupled microring resonators and directly measure the optical quality (Q) factor of the resonance modes. The Q factor is defined in terms of the the optical loss according to 
\begin{equation}
    \mathrm{\mathrm{Q} \equiv 2\pi \frac{energy\,\,stored}{energy\,\, lost\,\, per\,\, cycle} = \frac{\nu_\circ}{\Delta \nu_{FWHM}} }, 
\label{eq:Qdef}
\end{equation} 
where $\nu_\circ$ is the center frequency and $\mathrm{\Delta\nu_{\,FWHM}}$ is the full-width-at-half-maximum (FWHM) of the resonance. While our resonators are capable of supporting either transverse-electric (TE) or transverse-magnetic (TM) spatial modes, we choose to characterize the fundamental TE modes ($\mathrm{TE}_{00}$) whose electric field lie in the plane of the ring resonator and whose intensity more strongly samples the sidewall. Optical measurement of Q is performed via a laser frequency sweep across many TE modes of the resonators. To do this, we use a tunable continuous wave (CW) external cavity diode laser to probe the resonance modes between \SI{1510}{} and \SI{1630}{\nm}. The FWHM of the resonance is converted to frequency with a calibrated Mach-Zehnder interferometer (MZI).

The total, or loaded, quality factor as described in equation (\ref{eq:Qdef}) captures the optical energy loss per cycle due to a variety of contributions which we broadly classify as out-of-resonator/external (such as the bus-to-resonator coupling) and  intrinsic (such as intra-resonator absorption and scattering). It can be neatly expressed as an inverse sum of these two contributions,
\begin{equation}
    \mathrm{\Qloaded} = \left[\frac{1}{\mathrm{\Qext}} + \frac{1}{\mathrm{\Qint}}\right]^{-1}, 
\label{eq:Qloaded}
\end{equation}
where $\mathrm{\Qext}$ and $\mathrm{\Qint}$ are the external and intrinsic quality factors, respectively. $\mathrm{\Qext}$ and $\mathrm{\Qint}$ can be further stated in terms of their respective external and intrinsic loss rates, $\kappa_\circ$ and $\alpha_\circ$, as well as the speed of light and wavelength,
\begin{equation}
    \mathrm{\Qext} = \frac{2 \pi \,\mathrm{c}}{\lambda\,\kappa_\circ^2} \hspace{50pt}
    \mathrm{\Qint} = \frac{2 \pi \,\mathrm{c}}{\lambda\,\alpha_\circ^2} \,.
\end{equation}
Since we can directly measure $\mathrm{\Qloaded}$ but are interested in the loss contributions quantified by $\mathrm{\Qint}$, we vary the bus-to-resonator coupling gap and hence $\kappa_\circ$  to identify the condition where the external and intrinsic coupling rates are equivalent ($\mathrm{\Qint = \Qext}$). At this “critical coupling” point, the optical power coupled into the resonant modes is maximal and thus the on-resonance transmission of the out-coupled light is minimal. In this case, $\mathrm{\Qint}$ is simply twice $\mathrm{\Qloaded}$ while otherwise, $\mathrm{\Qint}$ is either,
\begin{equation}
    \mathrm{\Qint} = \frac{2\,\mathrm{\Qloaded}}{1-\sqrt\mathrm{{T}}}\quad(\mathrm{\text{overcoupled}}) \hspace{25pt} \mathrm{or} \hspace{25pt} 
    \mathrm{\Qint} = \frac{2\,\mathrm{\Qloaded}}{1+\sqrt{\mathrm{T}}} \quad(\mathrm{\text{undercoupled}}) ,
\label{eq:coupling}
\end{equation}
where $\mathrm{T}$ is the maximum transmission at the peak of the resonant mode.


\section{Results}

\subsection*{Quality Factors}

\begin{figure}[!htb]
\centering
\includegraphics[width=\linewidth]{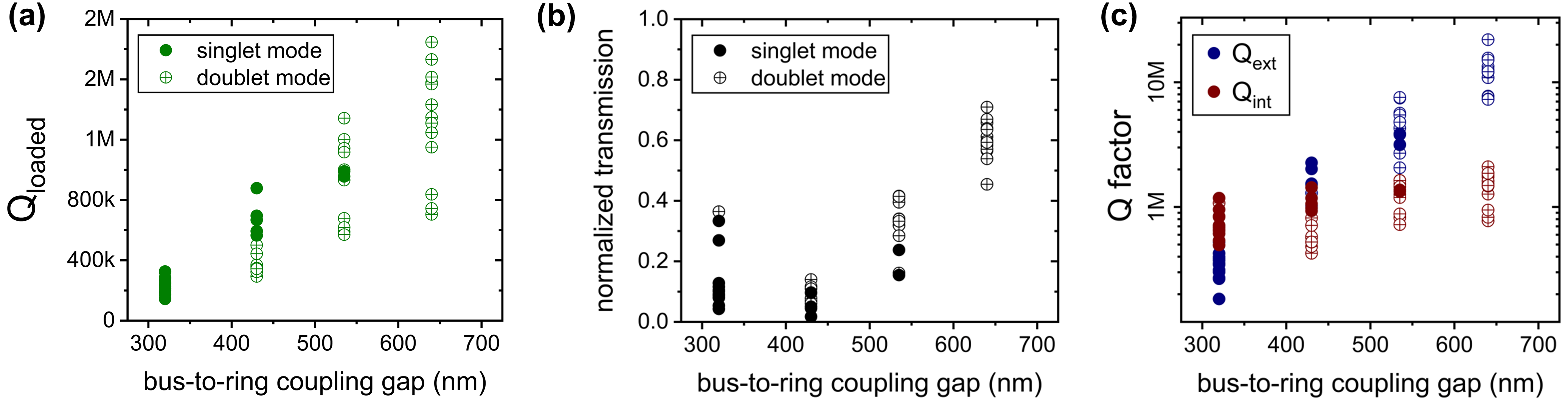}
\caption{Extracting optical quality factors for our resonators. (a) Loaded quality factor vs. coupling gap for resonance modes between \SI{1510}{} and \SI{1630}{\nm}. Split modes (doublets) are indicated by open circles with crosses. (b) Measured on-resonance transmission vs. gap. (c) $\mathrm{\Qext}$ and $\mathrm{\Qint}$ extracted from $\mathrm{\Qloaded}$ based on the coupling behavior in (b).}
\label{fig:Qbest}
\end{figure}

Figure \ref{fig:Qbest} demonstrates the intrinsic optical loss of typical resonators fabricated with our metal lift-off process. We fabricate a series of four waveguide-coupled resonators with sequentially increasing bus-to-ring coupling gaps between \SI{300}{} and \SI{700}{\nm} on a single chiplet. For enhanced mode sampling of the sidewall, resonators are designed with a small radius of \SI{23.3}{\micro\meter} corresponding to a free spectral range of \SI{1}{\tera\hertz}. The resist layer was patterned with a \SI{0.2}{\nA} beam current (\SI{\sim 5.8}{\nm} beam diameter) and a shot pitch of \SI{2}{\nm} (the smallest our machine will allow). The metal layer was deposited with an evaporation rate of \SI{0.3}{\angstrom\per\second} and a chamber base pressure of $4\mathrm{E}\minus7$. In Fig. \ref{fig:Qbest}a, we show $\mathrm{\Qloaded}$ as a function of the bus-to-ring coupling gap for both single peak (singlet) and split (doublet) fundamental TE resonance modes. We isolate $\mathrm{\Qext}$ and $\mathrm{\Qint}$ in Fig. \ref{fig:Qbest}c by identifying the critical coupling point from the transmission vs. coupling gap in Fig. \ref{fig:Qbest}b. While $\mathrm{\Qext}$ increases with gap, we expect $\mathrm{\Qint}$ to remain relatively constant, independent of coupling gap. We find $\mathrm{\langle\Qint\rangle = 900k \pm 70k}$ with a standard deviation of $\mathrm{290k}$ for the 19 singlet modes measured. 

\begin{figure}[!htb]
\centering
\includegraphics{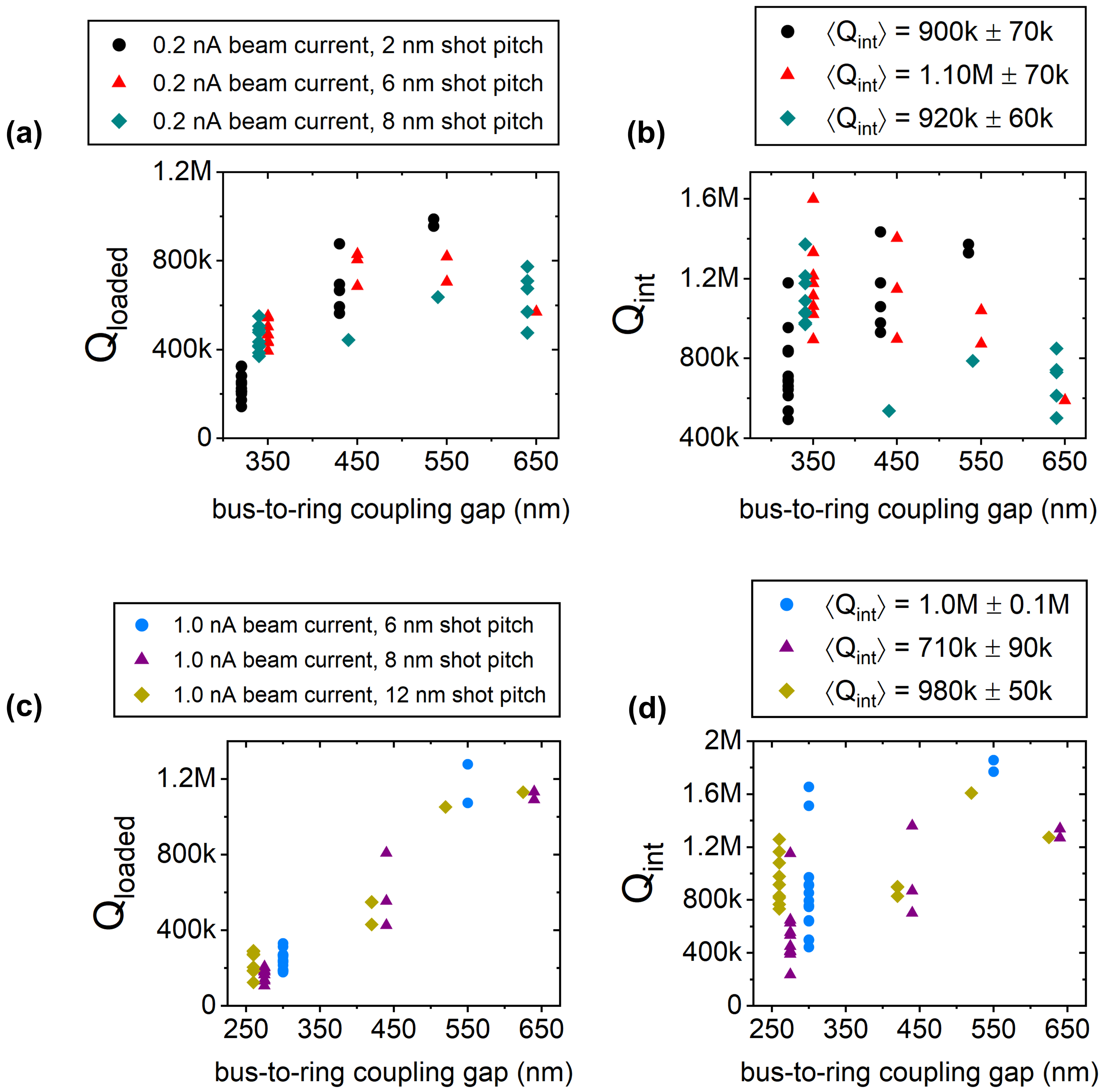}
\caption{(a) Loaded Q and (b) intrinsic Q vs. bus-to-ring coupling gap for resonance modes of rings patterned with \num{2}, \num{6} and \SI{8}{\nm} shot pitches at \SI{0.2}{\nA} beam current. (c) Loaded Q and (d) intrinsic Q for \num{6}and \SI{12}{\nm} shot pitches at \SI{1.0}{\nA} beam current.}
\label{fig:shotpitchcompare}
\end{figure}

We next consider the possible effect of the EBL parameters on the optical quality factor by fabricating the same series of rings, but with varying beam current and shot pitch. The results in Fig. \ref{fig:shotpitchcompare}a and b show the effect of increasing the shot pitch from \SI{2}{\nm} (Fig. \ref{fig:Qbest} results) to 6 and \SI{8}{\nm} with a common \SI{0.2}{\nA} beam current. The resonances in the \SI{6}{\nm} shot pitch devices yield $\mathrm{\langle\Qint\rangle = 1.10M \pm 70k}$ with a standard deviation of $\mathrm{250k}$, while the \SI{8}{\nm} shot pitch devices show a slightly lower $\mathrm{\langle\Qint\rangle = 920k \pm 60k}$ with a standard deviation of $\mathrm{250k}$; we consider both of these consistent with $\mathrm{\langle\Qint\rangle = 900k \pm 70k}$ for the \SI{2}{\nm} shot pitch devices. To remove any ambiguity, we only report singlet modes without visible doublet splitting.

We further explore the effect of shot pitch for a given beam size in Fig. \ref{fig:shotpitchcompare}c and d by increasing beam current to \SI{1.0}{\nA}, giving an increased beam diameter of \SI{\sim 7.5}{\nm}. At \SI{1.0}{\nA}, using a \SI{6}{\nm} shot pitch yields $\mathrm{\langle\Qint\rangle = 1.0M \pm 0.1M}$ with a standard deviation of $\mathrm{500k}$, while increasing the shot pitch to 8 and \SI{12}{\nm} results in $\mathrm{\langle\Qint\rangle = 710k \pm 90k}$ (standard deviation of $\mathrm{370k}$) and $\mathrm{\langle\mathrm{\Qint}\rangle = 980k \pm 50k}$ (standard deviation of $\mathrm{230k}$), respectively. We do not see evidence that the coarser shot pitch (and the significant difference between shot pitch and beam diameter for the \SI{1.0}{\nA}, \SI{12}{\nm} sample) leads to more sidewall loss. (Figures \ref{fig:shotpitchcompare}c and d show some potentially significant reduction in Q, strangely for the \SI{8}{\nm} shot pitch. However, this significance is small, and this data set is dominated by measurements in the smallest gap devices, where the coupling gap was measured to be close to the minimal critical dimension we can write. Due to this and due to the lack of a monotonic trend with shot pitch, we tentatively report the average intrinsic loss to be similar for all three conditions.) As with the results at \SI{0.2}{\nA}, the overall trend at \SI{1.0}{\nA} suggests that increasing the shot pitch has little effect on the average Q.

\begin{figure}[!htb]
\centering
\includegraphics{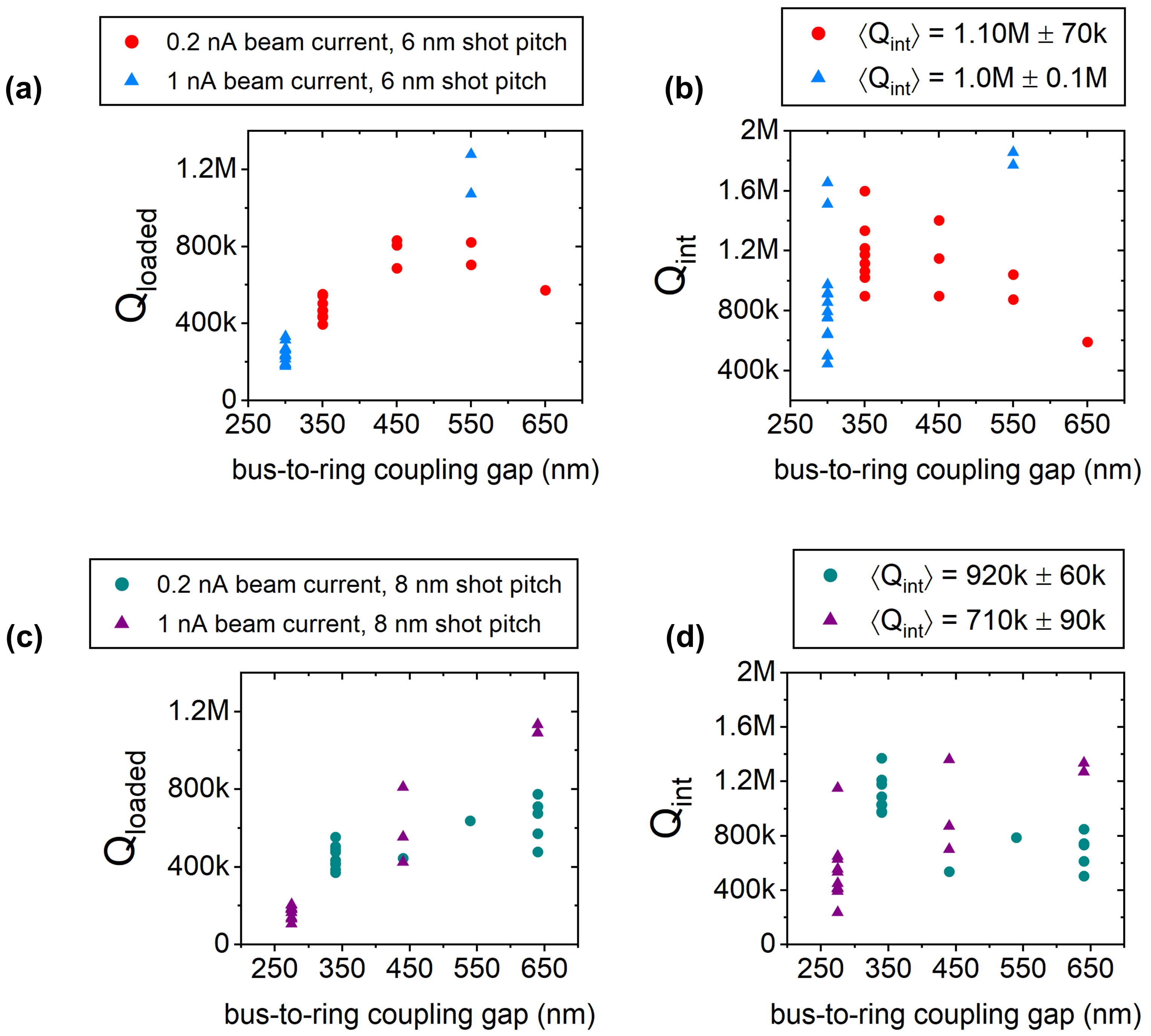}
\caption{(a) Loaded Q and (b) intrinsic Q vs. bus-to-ring coupling gap for resonance modes of rings patterned with a \SI{6}{\nm} shot pitch. (c) Loaded Q and (d) intrinsic Q for an \SI{8}{\nm} shot pitch at both \num{0.2} and \SI{1.0}{\nA} beam currents.}
\label{fig:beamcurrentcompare}
\end{figure}

We investigate the influence of beam current on optical loss by comparing the data in Fig. \ref{fig:shotpitchcompare} for varying beam currents and common shot pitch. As seen in Fig. \ref{fig:beamcurrentcompare}a and b for the \SI{6}{\nm} shot pitch, the \SI{0.2}{\nA} beam current results in $\mathrm{\langle\Qint\rangle = 1.10M \pm 70k}$ (standard deviation of $\mathrm{250k}$) while the \SI{1.0}{\nA} produces $\mathrm{\langle\Qint\rangle = 1.0M \pm 0.1M}$ (standard deviation of $\mathrm{500k}$). We find similar results holding \SI{8}{\nm} shot pitch constant (Fig. \ref{fig:beamcurrentcompare}c and d); \SI{0.2}{\nA} beam current results in $\mathrm{\langle\Qint\rangle = 920k \pm 60k}$ (standard deviation of $\mathrm{250k}$), and \SI{1.0}{\nA} beam yields $\mathrm{\langle\Qint\rangle = 710k \pm 90k}$ (standard deviation of $\mathrm{370k}$). Overall, the Q factors at either beam current suggest there is little influence of the beam current on the optical loss of our devices.

\begin{figure}[!htb]
\centering
\includegraphics{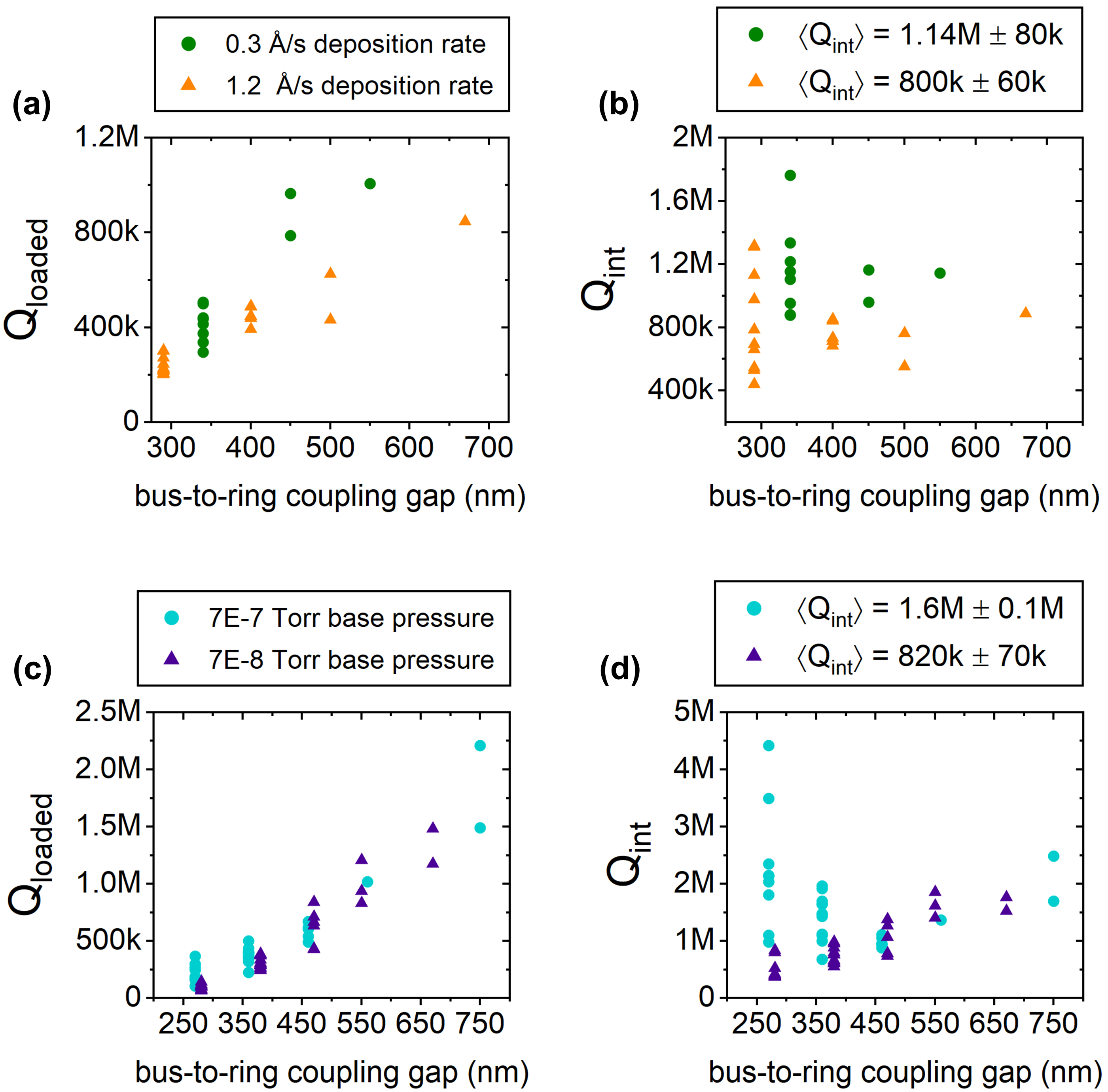}
\caption{(a) Loaded Q and (b) intrinsic Q vs. bus-to-ring coupling gap for resonance modes of rings with chromium mask deposited at \SI{0.3}{\angstrom\per\second} and \SI{1.2}{\angstrom\per\second} at 4E\minus7 base pressure. (c) Loaded Q and (d) intrinsic Q compare the results for rings patterned at \SI{0.5}{\angstrom\per\second} with 7E\minus7 and 7E\minus8 Torr base pressure.}
\label{fig:metaldepcompare}
\end{figure}

Finally, we consider the influence of the parameters used in depositing the metal mask layer starting with the metal deposition rate. We again pattern the same set of rings in EBL resist with our standard \SI{0.2}{\nA} beam current and \SI{2}{\nm} shot pitch and start by increasing the rate of the chromium mask deposition. The deposition rate is monitored in real time by a Quartz Crystal Microbalance (QCM) which provides a feedback signal for controlling the rate throughout the deposition process. Figures \ref{fig:metaldepcompare}a-b show the results of increasing the rate by a factor of 4 from our standard \SI{0.3}{\angstrom\per\second} to \SI{1.2}{\angstrom\per\second}. We observe that increasing the deposition rate results in a significant decrease in the intrinsic Q from $\mathrm{\langle\Qint\rangle = 1.14M \pm 80k}$ to $\mathrm{\langle\Qint\rangle = 800k \pm 60k}$ with standard deviations of $\mathrm{250k}$ and $\mathrm{240k}$, respectively. Of all the fabrication parameters investigated, we observe that the metal deposition rate has the clearest influence on Q. 

The metal deposition is performed in a high vacuum environment, with the starting base pressure controlled by the pump-down time allowed between loading and the start of deposition. Here, we explore two extremes of this pump-down time, 20 minutes (reaching 7E\minus7 Torr base pressure) and 14 hours (reaching 7E\minus8 Torr base pressure) before starting deposition. Since the chamber continues to pump down throughout the deposition process, we choose to use a slightly faster deposition rate of \SI{0.5}{\angstrom\per\second} to minimize the further reduction of the pressure throughout the \SI{\sim 20}{} minute process. As seen in Fig. \ref{fig:metaldepcompare}c and d, reducing the base pressure by a factor of ten results in a reduction of $\mathrm{\langle\Qint\rangle}$ by nearly half. We do not as of yet have a physical explanation for this trend. (It should be noted the statistics in this data are also dominated by devices with coupling gaps near our fabrication limit.) 


\subsection*{Resonance Splitting}

As indicated in Fig. \ref{fig:Qbest}, we observe a number of modes whose resonance peaks are split, which can hinder their use in applications requiring the large intracavity intensities provided by high-Q singlet microresonator modes. The origin of doublet resonance splitting is rooted in the fabrication-induced sidewall roughness of the microresonator waveguide \cite{Borselli2005, Morichetti2010}. The connection between the doublet splitting and degree of sidewall roughness originates in the strength of the back-scattered field and its coupling to the forward-propagating mode. This coupling strength is directly proportional to the back-scattered reflectivity, which itself is dependent on both the degree of sidewall roughness and the physical cavity length. Since our ring resonators are nominally the same dimensions in radius (as well as cross-section), we assume the physical path length is equivalent, and thus a measure of the resonance splitting gives us insight into the relative degree of fabrication-induced sidewall roughness.

Table \ref{tab:doublets} summarizes the measured doublet splitting as well as the prevalence (\% of all measured modes that are doublets) for the same samples used in the optical loss study. In Samples A-C, we consider the effects of varying shot pitch at \SI{0.2}{\nA} beam current. Increasing the shot pitch from \num{2} to \SI{8}{\nm}, we see a slight increase in the average doublet splitting from \SI{280\pm20}{\MHz} to \SI{320\pm20}{\MHz} suggesting a slight improvement in sidewall roughness for the \SI{2}{\nm} condition. 
In the \SI{1.0}{\nA} Samples D-F, we see a similar trend where increasing the shot pitch from \num{6} to \SI{12}{\nm} shows an increase in the splitting from \SI{230\pm20}{\MHz} to \SI{320\pm30}{\MHz} with little difference between the \num{8} and \SI{12}{\nm} devices. 
This suggests there may be an improvement in sidewall roughness, albeit minimal, for the finer shot pitch of \SI{6}{\nm}, potentially due to the slight overlap of adjacent shots.  

Next, we compare the effects of increasing beam current (and thus the beam size) at constant shot pitch, starting with Samples B (\SI{0.2}{\nA}) and D (\SI{1.0}{\nA}) at \SI{6}{\nm} shot pitch. There is a noticeable increase in the doublet splitting for the Sample B, where again the shot pitch is on-par with the beam diameter, compared to Sample D, where the beam diameter exceeds the shot pitch. 
Overall, based on the relative similarity of doublet splitting and prevalence across all of these samples, we conclude that the shot pitch and beam current parameters are not the main influence on sidewall roughness in our devices.

\begin{table}[!htb]
\newcolumntype{Y}{>{\centering\arraybackslash}X}
\newcolumntype{Z}{>{\hsize=0.75\hsize}Y}
\newcolumntype{U}{>{\hsize=1.25\hsize}Y}
\begin{tabularx}{\textwidth}{ | Z | Y | Y | Y | Y | Y | Y | Y | Y | U | }
\hline
\vspace{2pt}Sample\vspace{2pt} & \vspace{2pt}beam current\vspace{2pt} & \vspace{2pt}shot pitch\vspace{2pt} & \vspace{2pt}PVD rate\vspace{2pt} & \vspace{2pt}PVD base pressure\vspace{2pt} & \vspace{2pt}average doublet splitting\vspace{2pt} & \vspace{2pt}std. dev.\vspace{2pt} & \vspace{2pt}SE of mean\vspace{2pt} & \vspace{2pt}doublet: singlet ratio\vspace{2pt} & \vspace{2pt}prevalence\vspace{2pt} \\ 
\hline
\hline
& \vspace{3pt}nA\vspace{1pt} & \vspace{3pt}nm\vspace{1pt} & \vspace{3pt}\AA/s\vspace{1pt} & \vspace{3pt}Torr\vspace{1pt} & \vspace{3pt}MHz\vspace{1pt} & \vspace{3pt}MHz\vspace{1pt} & \vspace{3pt}MHz\vspace{1pt} & \vspace{3pt}-\vspace{1pt} & \vspace{3pt}\%\vspace{1pt}\\
\hline
\hline
\vspace{3pt} A & \cellcolor{red!25} \vspace{3pt} 0.2 & \cellcolor{red!25} \vspace{3pt} 2 & \vspace{3pt} 0.3 & \vspace{3pt} 4E-7 & \vspace{3pt} 280 & \vspace{3pt} 110 & \vspace{3pt} 20 & \vspace{3pt} 32:20 & \vspace{3pt} 62 \\
B& \cellcolor{red!25} 0.2& \cellcolor{red!25} 6 & 0.3& 4E-7& 300& 140& 20& 41:14& 75\\
C& \cellcolor{red!25} 0.2& \cellcolor{red!25} 8 & 0.3& 4E-7& 320& 120& 20& 35:21& 63\\
D& \cellcolor{orange!25} 1.0 & \cellcolor{orange!25} 6 & 0.3& 4E-7& 230& 90& 20& 26:16& 62\\
E& \cellcolor{orange!25} 1.0 & \cellcolor{orange!25} 8 & 0.3& 4E-7& 300& 170& 40& 25:17& 60\\
F& \cellcolor{orange!25 } 1.0 & \cellcolor{orange!25} 12 & 0.3& 4E-7& 320& 150& 30& 39:19& 65\\
G& 0.2& 2& 0.5& \cellcolor{blue!25} 7E-7 & 260& 170& 30& 33:30& 52\\
H& 0.2& 2& 0.5& \cellcolor{blue!25} 7E-8 & 220& 80& 10& 33:36& 48\\
I& 0.2& 2& \cellcolor{green!25}0.3& 4E-7 & 240& 110& 20& 20:11& 65\\
J& 0.2& 2& \cellcolor{green!25}1.2& 4E-7& 390& 170& 30& 29:20& 60\\
\hline
\end{tabularx}
\caption{Table of EBL and metal deposition parameters and associated doublet splitting with samples labeled alphabetically and color coded according to their varied parameters. Slight differences in the number of modes used in the averaging vs the doublet:singlet ratio exist as some modes were recorded as doublets, but the splitting was minimal and peak centers could not be determined reliably. These modes were still recorded as doublets due to the observable, but unmeasurable splitting.}
\label{tab:doublets}

\end{table}

Shifting our focus to the next major process in the fabrication flow, we consider the influence of the parameters of the metal deposition process. Samples G and H compare the results for variation in starting base pressure; Sample G (7E\minus7 Torr base pressure) was measured to have an average doublet splitting of \SI{260\pm30}{\MHz}, a slight increase over the \SI{220\pm10}{\MHz} splitting in Sample H (7E\minus8 Torr), while the doublet prevalence is essentially the same. These results suggest that samples patterned at the lower base pressure have a slightly lower degree of sidewall roughness; this may disagree with the results of Fig. \ref{fig:metaldepcompare}d, which may further indicate a fabrication issue with gaps much smaller than \SI{300}{\nm}. (For completeness, we report observation of defects from the metal lift-off in Sample H, mainly in the coupling region of the \SI{\sim 300}{} and \SI{\sim 400}{\nm} gap devices, while Sample G was clear of defects. Despite these defects, there is no major effect on the number of doublets in the \SI{\sim 300}{} and \SI{\sim 400}{\nm} gap devices. For more details on common defects associated with metal lift-off, refer to the Supplementary Information document.)

Finally, we explore the influence of the metal deposition rate in Samples I and J where the deposition rate is increased respectively from \num{0.3} to \SI{1.2}{\angstrom\per\second} at a comparable starting base pressure. Increasing the deposition rate results in an increase in the average doublet splitting from \SI{240\pm20}{\MHz} to \SI{390\pm30}{\MHz}, while the doublet prevalence is comparable for both devices. Additionally, with both samples free of the lift-off defects observed in other samples, we take this large increase in the average splitting to indicate that the deposition rate is the main driver of roughness in the metallic mask and also the sidewall roughness transferred from the mask during the etch process. This is consistent with our findings linking metal deposition rate to optical loss.


\subsection*{Frequency Comb Generation}

\begin{figure}[!htb]
\centering
\includegraphics[width=\linewidth]{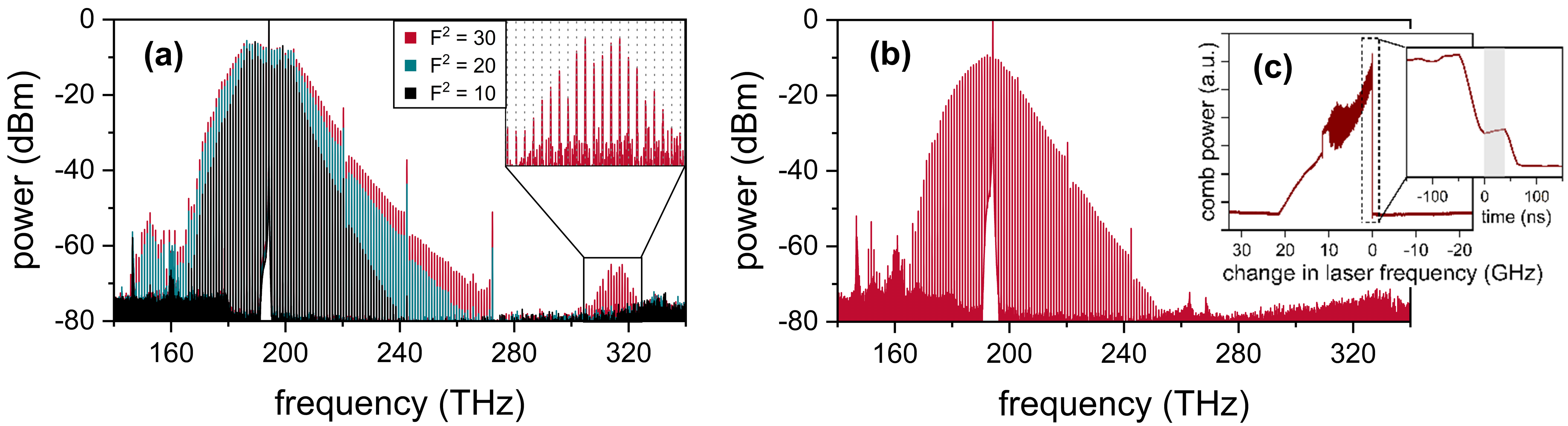}
\caption{(a) Comb spectra for three different optical pump powers: $\mathrm{F^2=10}$ (black), $\mathrm{F^2=20}$ (green), and $\mathrm{F^2=30}$ (red). The emergence of dual dispersive waves extend the spectral bandwidth over an octave. The inset zooms in on the short-wavelength dispersive wave and compares it to a grid of the comb mode spacing extrapolated and continuously extended from the central comb modes (gray dashed lines). (b) Optical soliton spectrum for a long-lived soliton obtained at $\mathrm{F^2=30}$, using the pump frequency sweeping method to stabilize the soliton state after generation. (c) The short-lived soliton step, lasting for roughly \SI{40}{\ns}, observed in the total comb power during an adiabatically slow laser frequency sweep across the resonance. (This is the same soliton state stabilized in (b).)}
\label{fig:MI comb and soliton}
\end{figure}

Intrinsic quality factors near 1 million in our THz-rate resonators suggest that they are sufficient to support broadband frequency comb generation\cite{Li2017}. Optically pumping a high-Q, critically coupled mode of Sample A, we observe a threshold for nonlinear four-wave mixing of \SI{6.5}{\mW} in the bus waveguide and generate optical frequency combs at three different multiples of this threshold power (Fig. \ref{fig:MI comb and soliton}a). Here the CW pump power in the waveguide is reported in units of the threshold power, $\mathrm{F^2\equiv P_{pump}/P_{threshold}}$. These three comb spectra represent low-coherence chaotic states where the comb is formed as the result of cascaded four-wave mixing in the resonator. As CW pump power increases, the comb power and the bandwidth increase. We purposefully design our resonator dimensions to achieve dual dispersive wave, octave-bandwidth frequency combs, a feat which requires precise control over the fabricated dimensions and geometry. For a pump powers of $\mathrm{F^2=20}$ and $\mathrm{F^2=30}$, we observe a dispersive wave centered around \SI{152}{\THz} (\SI{1965}{\nm}). At the largest pump power ($\mathrm{F^2=30}$), we observe the emergence of a second dispersive wave near \SI{315}{\THz} (\SI{950}{\nm}). Together, these dispersive waves extend the overall bandwidth of our comb to over an octave. 

While we find chaotic combs helpful indicators of resonator dispersion, many applications benefit from the high coherence and smooth spectral envelope of soliton pulsed operation\cite{Drake2019}. 
The transformation from chaos to soliton can be observed as a sharp drop in the intracavity power to a steady level when sweeping the pump frequency across the resonant mode\cite{Briles2020}. The transition to soliton operation and associated reduction in power is illustrated Fig. \ref{fig:MI comb and soliton}c, which shows the intracavity power during a slow adiabatic sweep (roughly \SI{1}{\GHz\per\micro\second} at a moderate optical power of $\mathrm{F^2=12}$ in the waveguide) of the laser frequency across the resonance. We observe a build-up of the comb power as the intracavity field develops through the noisy chaotic comb regime on the rising edge. The comb power reaches a maximum at which point the laser crosses the peak of the resonance and falls into the soliton regime indicated by a short step of \SI{40}{\ns} duration (highlighted gray region in the inset). Stabilizing the soliton state for extended operation requires compensating for the dramatic thermal shifts involved with the chaotic comb-soliton transition. We achieve this with a fast pump frequency sweep method that allows us to sweep multiple orders of magnitude faster than the piezo modulation limit of our laser\cite{Stone2024}.

Utilizing this method, we successfully create a long-lived soliton pulse in our devices, with resulting spectrum shown in Fig. \ref{fig:MI comb and soliton}b for a $\mathrm{F^2=30}$ pump. Although our microrings were designed to support octave-bandwidth spectra, we did not optimize the bus waveguide-resonator coupler for broadband extraction, instead opting for a point coupler (straight bus waveguide) which allowed for simpler resonator mode analysis. We believe that with improved coupler design, we can generate and extract soliton combs with dual dispersive waves at sufficient power for f-2f self referencing\cite{Moille2019,Tan2010,Briles2018,Drake2019}.


\section{Conclusions and Discussion}

In this work, we have demonstrated that subtractive fabrication utilizing an etch mask formed by metal lift-off is a viable technique for realizing photonic waveguides in thick, etch-resistant dielectric films. With this method we achieve $\mathrm{\Qint}$ values close to 1 million in THz-rate ring resonators with near-vertical sidewalls and well defined high-aspect-ratio features. Through careful investigation of our fabrication process parameters, we identify the metal deposition, specifically the deposition rate, is currently the predominant factor influencing our sidewall loss, more so than electron beam lithography parameters. We believe that adjusting the metal choice and deposition details will result in further reduction in optical loss. 

As the field of photonics continues to expand, new thresholds for performance, widened scope in applications, and advances in integration will place ever-increasing demands on the fabrication of photonic devices. To meet these demands, robust solutions to fabricating complex, high-aspect-ratio designs in a variety of hard-to-etch photonic materials will be of great use. In addition, the progressive efforts to translate photonics research from the lab to integrated devices in the field and commercial sphere require techniques that can be applied by anyone from the novice researcher working in a small research cleanroom to experienced technicians in high-throughput commercial foundries. The results of this work highlight that metallic etch masks formed by metal lift-off can serve this purpose as a widely applicable and easy-to-implement technique that meets all of these requirements. With just a few steps added to the typical fabrication flow, minimal optimization, and the use of standard tools found in most CMOS foundries, this method is capable of producing high quality photonic devices such as the high-Q microresonators presented in this work. With further improvement, we speculate that microresonators formed with this method could reach the quality factors attainable with more complex techniques, with fewer trade-offs and limitations in device design. 


\section{Supplementary Information}

\noindent Further information regarding potential process defects can be found in the Supplementary Information document.


\bibliography{references}


\section*{Acknowledgments}

We gratefully acknowledge financial support from AFOSR (FA9550-22-1-0174), NSF (2410813), NSF EPSCoR (2217786), and UNM Women in STEM awards. We thank the scientists at Sandia National Labs' Center for Integrated Nanotechnologies (CINT) for their expertise and guidance, with particular thanks to John Nogan, Anthony James, and Michael De La Garza for their invaluable insight on process development. Finally, we thank Joseph Yelk for sharing his expertise in scientific illustration and for his contributions to Fig. \ref{fig:fabrication flow}.


\section*{Data Availability}

The data that support the findings of this study are available from the corresponding author upon reasonable request.


\section*{Author contributions statement}

The contributions of the authors to this research are as follows: G.M.C. conceptualized this study, developed the fabrication process, performed fabrication and characterization experiments, analyzed the results, and wrote the manuscript. L.R. and F.B. performed characterization experiments and data analysis. L.R. contributed to process development and study conceptualization. T.E.D. conceptualized this work and provided guidance, training, and oversight. All authors reviewed the manuscript. 


\section*{Additional information}

The authors declare no competing interests.


\end{document}


\maketitle

\section{Redeposition, incomplete patterning, and excess metal buildup: characteristic issues in metal lift-off and their signatures in the waveguides}

\renewcommand{\thefigure}{S\arabic{figure}}
\renewcommand{\thetable}{S\arabic{table}}
\renewcommand{\theequation}{S\arabic{equation}}

\begin{figure}[h!]
    \centering
    \includegraphics[scale=0.75]{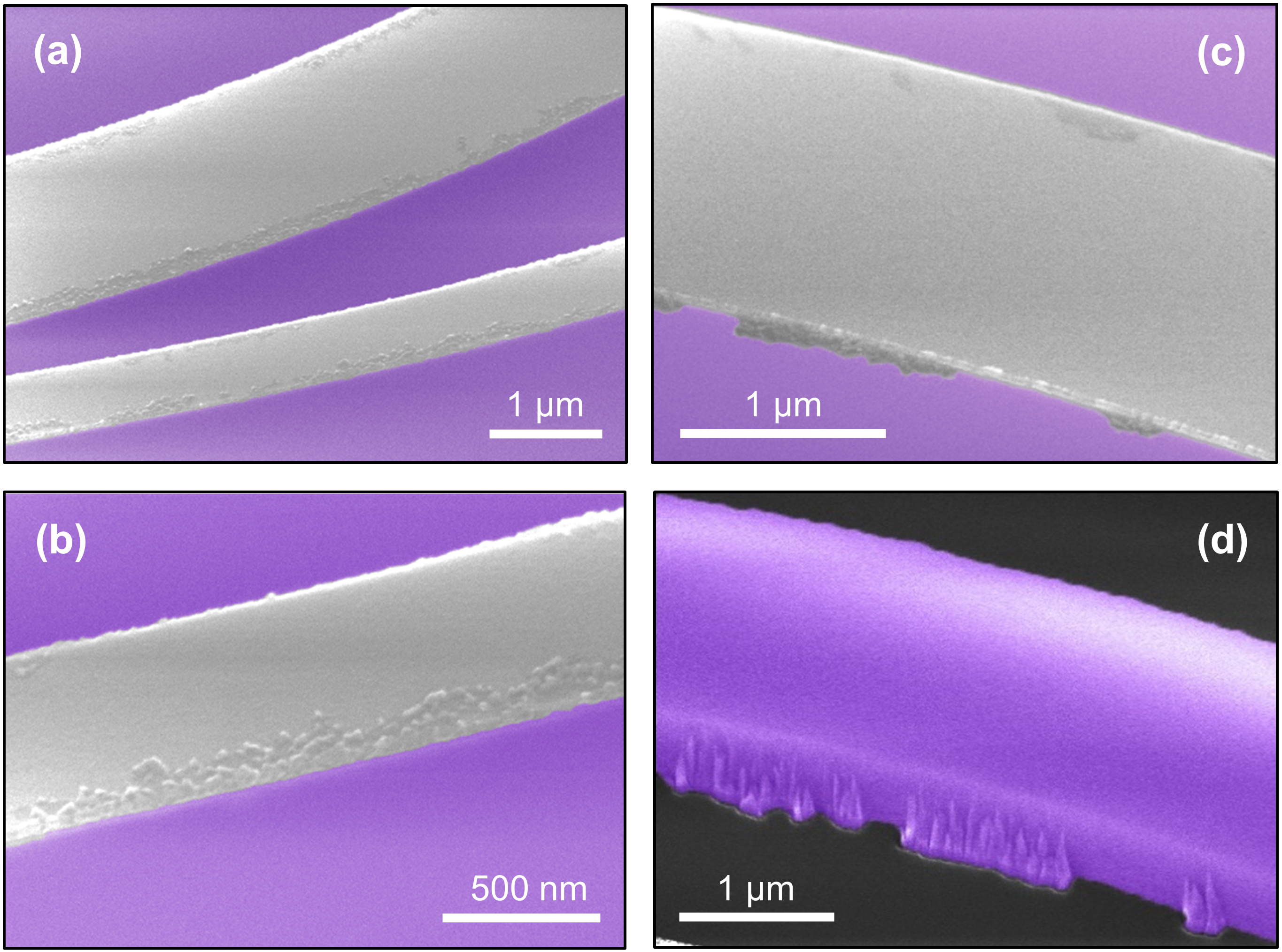}
    \caption{SEM micrographs of lift-off defects and their effects post-etching with false coloring for silicon nitride (purple) and chromium (silver). (a) SEM micrograph showing impartial lift-off with (b) zoomed focus on the extra metal and roughness along the perimeter of the metal mask for the bus (lower) waveguide. The metal mask is shown on the bare silicon nitride prior to etching. (c) Lift-off defects extending past sidewall perimeter due to either impartial lift-off or metal re-deposition resulting in unwanted masking. The metal mask is shown on the bare silicon nitride prior to etching. (d) Partially etched "islands" due to the defect masking in (c).}
    \label{fig:defects}
\end{figure}

While metal deposition and lift-off has the advantage of being a relatively simple addition to the subtractive fabrication flow, it introduces additional defect-prone steps that can carry flaws through to the final device in the subsequent etch. Aside from the quality of the metal layer (which determines the results in Fig. 6), metal deposition on the sacrificial resist sidewall, incomplete/partial lift-off, and re-deposition of the floated metal during lift-off are all potential pitfalls that must be considered and avoided. We highlight these three issues in Fig. \ref{fig:defects}. Partial lift-off due to sidewall deposition (Fig. \ref{fig:defects}a and \ref{fig:defects}b (zoomed)) can result in post-etch roughness transferred to the sidewall after etching. In cases where the defects extend past the sidewall, or there is re-deposition of the lifted metal in the open areas (Fig. \ref{fig:defects}c), defects such as the partially etched "islands" along the waveguide sidewalls (seen in Fig. \ref{fig:defects}d) may be observed. For the issue of sidewall deposition resulting in poor mask edge quality and/or poor lift-off, targeting an "undercut" profile in the resist sidewall can aid in shielding the sidewall from unwanted metal deposition. If re-deposition in the subsequent lift-off stage is an issue, heating the solvent and/or performing the lift-off in an ultrasonic bath may help remove the lifted metal from the surface and away from the open area to be etched. 
We have found that SEM imaging can clearly identify the above defects associated with metal lift-off, providing a clear way to immediately identify workable devices.